# A method to convert traditional fingerprint ACE / ACE-V outputs ("identification", "inconclusive", "exclusion") to Bayes factors

**Author and affiliations:**

Geoffrey Stewart Morrison [1,2,*]

[1] Forensic Data Science Laboratory, Aston University, Birmingham, UK

[2] Forensic Evaluation Ltd, Birmingham, UK

*Corresponding author: G.S. Morrison, e-mail: geoff-morrison@forensic-evaluation.net

**ORCID:**

Geoffrey Stewart Morrison        0000-0001-8608-8207

**Disclaimer:**

All opinions expressed in the present paper are those of the author, and, unless explicitly stated otherwise, should not be construed as representing the policies or positions of any organizations with which the author is associated.

**Declaration of competing interest:**

The author declares that they have no known competing financial interests or personal relationships that could have appeared to influence the work reported in this paper.

**Acknowledgements:**

This research was supported by Research England's Expanding Excellence in England Fund as part of funding for the Aston Institute for Forensic Linguistics 2019–2023.

My thanks to William Morris and to Lisa J. Hall for comments on earlier versions of this paper. Any inadequacies are solely my responsibility.

**Highlights**

- Fingerprint examination conclusions

- Conversion of "identification", "inconclusive", "exclusion" to Bayes factors

- Beta-binomial model

- Uninformative priors

- Informative priors

# A method to convert traditional fingerprint ACE / ACE-V outputs ("identification", "inconclusive", "exclusion") to Bayes factors

## Abstract

Fingerprint examiners appear to be reluctant to adopt probabilistic reasoning, statistical models, and empirical validation. The rate of adoption of the likelihood-ratio framework by fingerprint practitioners appears to be near zero. A factor in the reluctance to adopt the likelihood-ratio framework may be a perception that it would require a radical change in practice. The present paper proposes a small step that would require minimal changes to current practice. It proposes and demonstrates a method to convert traditional fingerprint-examination outputs ("identification", "inconclusive", and "exclusion") to well-calibrated Bayes factors. The method makes use of a beta-binomial model, and both uninformative and informative priors.

## Abbreviations

ACE – analysis, comparison, and evaluation

ACE-V – analysis, comparison, evaluation, and verification

ASB – Academy Standards Board (American Academy of Forensic Sciences)

B – Bayes factor

$c$ – count

d – different source

DFSC – Defense Forensic Science Center of the United States Department of the Army

ENFSI – European Network of Forensic Science Institutes

EX – exclusion

ID – identification

IN – inconclusive

Λ – likelihood ratio

$m$ – pseudo number of fingermark-fingerprint pairs

$n$ – number of fingermark-fingerprint pairs

θ – probability

*RS* – response

s – same source

*t* – truth

# 1  Introduction

Fingerprint examiners appear to be reluctant to adopt probabilistic reasoning, statistical models, and empirical validation (Cole [1], [2]; Mnookin et al. [3]; Curran [4]; Morrison & Stoel [5]; Swofford et al. [6]). The rate of adoption of the likelihood-ratio framework by fingerprint practitioners appears to be near zero (Bali et al. [7]; Cole & Barno [8]). A factor in the reluctance to adopt the likelihood-ratio framework may be a perception that it would require a radical change in practice. The present paper makes a proposal that would require minimal changes to current practice. It proposes a method to convert traditional fingerprint-examination outputs to well-calibrated Bayes factors.[1]

In current fingerprint-examination practice, conclusions are most commonly reported as "identification" (or "individualization"), "exclusion", or "inconclusive" (Expert Working Group on Human Factors in Latent Print Analysis [9]; Cole [2]; Thompson et al. [10]; Forensic Science Regulator [11]). Traditionally, "identification" corresponds to a posterior probability of 1 and "exclusion" to a posterior probability of 0, with "inconclusive" as a no-conclusion option rather than a probabilistic value between 0 and 1.

Proposals have been made that keep the terms "identification" and "exclusion", but state them as the examiner's opinion, rather than as facts, and redefine them to mean probabilities very very close to but not exactly 1 and 0, e.g., United States Department of Justice [12]: "'Source identification' is an examiner's conclusion that two friction ridge skin impressions originated from the same source. ... A 'source identification' is the statement of an examiner's opinion ... that the probability that the two impressions were made by different sources is so small that it is negligible." This approach has been criticized in Expert Working Group on Human Factors in Latent Print Analysis [9] pp. 72–73, Cole [2], and Thompson et al. [10] pp. 60–62. The difference between the original and revised definitions is negligible, and, without a change in nomenclature, triers of fact and others are likely to continue interpreting "identification" and "exclusion" on face value, i.e., as probabilities of 1 and 0. Knowing that a fingerprint examiner's opinion is that the mark and print came from the same source is of little value unless one knows the probability that the practitioner would opine that the mark and print came from the same source if they really did come from the same source versus the probability that the practitioner would opine that the mark and print came from the same source if they actually came from different sources (President's Council of Advisors on Science and Technology [13]; Morrison et al. [14]).

---

[1] Bayes factors are the Bayesian analogues of likelihood ratios.

Proposals have also been made to move away from the three-level ("identification", "inconclusive", "exclusion") opinion scale and adopt ordinal opinion scales with more levels, e.g., in the most recent publicly released (November 2021) draft of the ASB 013 Standard for Friction Ridge Examination Conclusions.[2] That draft claims to have a 5-level opinion scale, but it actually has 9 levels. The levels are labelled: "source identification", "inconclusive with similarities", "inconclusive", "inconclusive with dissimilarities", and "source exclusion", but each of "inconclusive with similarities" and "inconclusive with dissimilarities" is further divided into "weak", "moderate", and "strong". The levels of the draft opinion scale are associated with verbal expressions of degree of support for the same-source hypothesis relative to degree of support for the different-source hypothesis, e.g., "the observed data provide more support for the proposition that the impressions originated from different sources rather than the same source". On their face, these "support" statements appear to be expressions of posterior odds, but additional wording, e.g., "the examiner believes the observed data are more probable if the impressions have different sources than the same source" suggests that they are intended to be verbal expressions of likelihood ratios. The highest and lowest levels of the opinion scale are, however, still labelled "identification" and "exclusion". How an examiner is to evaluate strength of evidence in a way that would lead to the selection of the appropriate level on the opinion scale is vague: "An examiner may utilize their knowledge, training, and experience as well as a statistical model".[3] The ASB 013 draft has flaws, but it is clearly an attempt to move away from only stating conclusions that are qualitative expressions of posterior probabilities that are (or are very very close to) either 1 or 0. It is too early to tell whether there will be major changes between the current draft and the final version of ASB 013, or whether the final version will be widely adopted by examiners.

---

[2] https://www.aafs.org/sites/default/files/media/documents/013_Std_Ballot02.pdf

[3] In a standard, a sentence with "may" gives permission. This sentence therefore states what examiners are permitted to do, not what they are recommended or required to do.

The European Network of Forensic Science Institutes (ENFSI) Guideline for Evaluative Reporting in Forensic Science [15] recommends that forensic practitioners subjectively assign a number between 0 and 1 for the numerator of a likelihood ratio, subjectively assign a number between 0 and 1 for the denominator, then divide the former by the latter. The ENFSI Guideline recommends that forensic practitioners report the subjectively assigned numerical likelihood-ratio value and/or a corresponding verbal expression from an ordinal opinion scale. Each level on the opinion scale is associated with a range of numerical likelihood-ratio values, and each level has an associated verbal expression of relative degrees of support for the hypotheses and an associated verbal expression of a likelihood ratio. For example, the numerical likelihood-ratio range 100–1000 is associated with the following verbal expressions: "The forensic findings provide moderately strong support for the first proposition relative to the alternative." "The forensic findings are appreciably more probable given one proposition relative to the other." The ENFSI Guideline recommends that a numerical likelihood-ratio value be subjectively assigned first and that it then be converted to a verbal expression from a level on the ordinal opinion scale, not the other way around.[4] The recommendations of the ENFSI Guideline do not appear to have been widely adopted by fingerprint examiners. Reporting of uncalibrated and unvalidated subjective assignment of likelihood-ratio values has been criticized in Thompson et al. [10] p. 65 and in Morrison et al. [18].

In 2017, the Defense Forensic Science Center (DFSC) of the United States Department of the Army proposed that fingerprint examiners state their conclusions as subjectively-assigned numerical likelihood-ratio values based on degree of correspondence between the questioned-source fingermark and the known-source fingerprint: "The probability of observing this amount of [corresponding ridge detail] is approximately ## times greater when impressions are made by the same source rather than by different sources."[5] In Swofford et al. [19], members of DFSC (in collaboration with others) also proposed "FRStat", a method for providing "statistical assessment of the strength of fingerprint evidence" based on similarity scores calculated from comparisons of minutiae annotations. FRStat has a passing resemblance to methods for calculating likelihood-ratio values, but it calculates tail probabilities for similarity scores, not likelihood ratios. The flaws with this approach are comprehensively described in Neumann [20]. Results from FRStat and reporting using DFSC's wording have been tendered as evidence in US military courts and in at least one civilian case (Neumann [20]; Swofford et al. [21]).

---

[4] If, instead of subjective assignment of a likelihood-ratio value, a likelihood-ratio value is calculated using relevant data, quantitative measurement, and statistical models, Marquis et al. [16], quoting Berger et al. [17], recommend that only the calculated number be reported.

Part of the reluctance to adopt the likelihood-ratio framework for evaluation of forensic evidence may be because of the perception that it would require a radical change in practice. The present paper proposes a small step that would require minimal changes to current practice. In this proposal, fingerprint examiners continue with their existing practice and state ACE or ACE-V outputs as "identification", "inconclusive", or "exclusion". Those outputs are subsequently converted to well-calibrated Bayes factors using a statistical model. The model is trained using data which consist of fingerprint examiners' "identification", "inconclusive", and "exclusion" responses to fingermark-fingerprint pairs for which the true same-source or different-source status in known. The statistical model could be applied to the output of an ACE process conducted by a single fingerprint examiner, or could be applied to the output of an ACE-V process to which two fingerprint examiners contribute. The present paper demonstrates use of the statistical model for the ACE output of individual examiners. A separate model is trained for each fingerprint examiner. The model is therefore calibrated to reflect the strength of evidence associated with that fingerprint examiner stating each of the three

---

[5] Quoted from Neumann [20].

outputs.

The scope of the present paper is modest. It describes and demonstrates a statistical model as a proof of concept, and it discusses some considerations with respect to what would be needed to transition the method into practice.

## 2 Methodology

### 2.1 Data

The data used for the proof of concept are taken from Langenburg et al. [22]. Participants gave "identification", "inconclusive", or "exclusion" responses to each of 12 fingermark-fingerprint pairs (7 same-source pairs and 5 different-source pairs). The fingermark-fingerprint pairs were selected to be challenging.

Each participant was assigned to one of six groups. For the present study, we make use of data from participants in Group 1, the control group, who performed their examination as usual without being supplied with additional "tools". For the present paper, data from participants who were not practicing fingerprint examiners have been excluded, leaving data from 24 participants.

For the purposes of the present paper, it is assumed that the fingermark-fingerprint pairs in Langenburg et al. [22] all represented the same set of conditions, hence a statistical model trained using these data can be generalized for use with other fingermark-fingerprint pairs that also have that set of conditions. The conditions for a case involve the quality of the fingermark and the quality of the fingerprint, but if fingerprints are high-quality it is the quality of the fingermark that will be key in defining the conditions for the case. Deciding whether data used for training (including calibration) and for validation are sufficiently reflective of the conditions of a case requires a subjective judgement which requires subject-area expertise, see the Consensus on Validation of Forensic Voice Comparison (Morrison et al. [23]).

## 2.2 A likelihood-ratio model

Table 1 lists the symbols that that will be used to represent counts of each potential response to each truth value, i.e., counts of "identification", "inconclusive", or "exclusion" in response to whether fingermark-fingerprint pairs were same-source or different-source pairs. The symbol $c$ represents a count, subscripts s and d represent the truth as to whether the pair was a same-source pair or a different-source pair respectively, and subscripts ID, IN, and EX represent whether an examiner's response was "identification", "inconclusive", or "exclusion" respectively. $n_s$ and $n_d$ represent the number of same-source pairs and different-source pairs respectively. In the Langenburg et al. (2012) data $n_s = 7$ and $n_d = 5$.

**Table 1.** Symbols for counts of "identification", "inconclusive", or "exclusion" outputs in response to whether fingermark-fingerprint pairs were same-source or different-source pairs.

|  |  | Response |  |  | Number of pairs |
|---|---|---|---|---|---|
|  |  | identification | inconclusive | exclusion |  |
| **Truth** | same source | $c_{(ID|s)}$ | $c_{(IN|s)}$ | $c_{(EX|s)}$ | $n_s$ |
|  | different source | $c_{(ID|d)}$ | $c_{(IN|d)}$ | $c_{(EX|d)}$ | $n_d$ |

Given the response counts for an examiner, a likelihood-ratio value associated with each response category could be calculated as in Equation (1), in which $\Lambda$ represents a likelihood ratio, $\hat{\theta}$ is an estimated probability value, and subscript $RS$ (response) is a placeholder for ID, IN, or EX ($RS = \{ID, IN, EX\}$). The likelihood-ratio value is calculated as the proportion of responses that are a particular response when the pair is a same-source pair divided by the proportion of responses that are that particular response when the pair is a different-source pair.

(1)

$$\Lambda_{RS} = \frac{\hat{\theta}_{(RS|\text{s})}}{\hat{\theta}_{(RS|\text{d})}} = \frac{c_{(RS|\text{s})}/n_\text{s}}{c_{(RS|\text{d})}/n_\text{d}}$$

The responses are considered a sample of the population of potential responses, i.e., a population defined as all the responses that the examiner could potentially give to all fingermark-fingerprint pairs that have the same set of conditions (see §2.1). The sample is used to provide an estimate for what the likelihood-ratio value would be if one were able to calculate it using the entire population of data, i.e., it is an estimate of the examiner's theoretical underlying "true" performance under the tested conditions. A problem occurs, however, when the amount of sample data is small. For example, if an examiner responded "identification" to 1000 out of 10,000 different-source pairs then one would be confident that that practitioner's "true" false-alarm rate was very close to 10%. If an examiner responded "identification" to 1 out of 10 different-source pairs then one's best estimate for that practitioner's "true" false-alarm rate would be 10%, but one would have a lot of uncertainty about how close that estimate was to that examiner's "true" false-alarm rate. Another problem which occurs with small sample sizes is the high probability of obtaining a zero count, e.g., if the "true" false-alarm rate were 1% and the sample size were 10, then the probability of obtaining a zero count from a sample would be high. If there were a zero count in the numerator in Equation (1) then the calculated likelihood-ratio value would be 0, and if there were a zero count in the denominator then the calculated likelihood-ratio value would be infinite. A solution to these problems is to adopt a Bayesian approach and calculate a Bayes factor instead of a likelihood ratio (see §2.3).

## 2.3 A Bayes-factor model

Philosophically, in a frequentist approach, one attempts to calculate a probability or likelihood value that is an estimate of a true but unknown value. In contrast, in a Bayesian approach, probabilities and likelihoods are states of belief. A Bayesian begins with a belief about the value of a statistical parameter of interest, observes sample data, and based on the sample data they update their belief about the value of that parameter. A rigorous Bayesian will justify their prior belief and will have prespecified the model that they will use for representing and updating their belief (Jaynes [24] p. 373). If others accept the justification for the prior and choice of model as reasonable then they should also be willing to adopt for themselves the posterior belief about the parameter value. The posterior is the result of a mixture of the prior and the sample data. If the amount of sample data is large, the posterior depends heavily on the sample data, but if the amount of sample data is small, the weight that the sample data contribute to the posterior is less. If the amount of sample data is small, the weight contributed by the prior is higher relative to the weight it contributes if the amount of sample data is large. This provides a solution to the problems described above with respect to small sample sizes. The priors, however, must be chosen and justified.

Priors can be "informative" or "uninformative". Informative priors can be based on existing relevant information. For example, if the performance of a fingerprint examiner has not been previously tested under the conditions of interest, then a reasonable informative prior could be based on the assumption that this examiner's performance is the same as the average of that of all examiners who have already been tested under these particular conditions. Alternatively, if the performance of a fingerprint examiner has not been previously tested under the conditions of interest but the examiner has been tested under somewhat similar conditions, then a reasonable informative prior could be based on the assumption that the examiner's performance on these particular conditions will be the same their performance on the somewhat similar conditions under which they have already been tested. If no relevant information is available, then an uninformative prior would be a reasonable choice. In may be argued that no prior is completely uninformative, but there are relatively uninformative "reference" priors (e.g., Jeffreys' reference priors) whose use is uncontroversial (Jeffreys [25]; Jaynes [26]; Bernardo [27]; Berger et al. [28]).

The proposed method is outlined in Figure 1. For each of the numerator and the denominator of the Bayes factor, the proposed method first uses a model with uninformative prior hyperparameter values, then updates the model using the sample data from an examiner and thereby arrives at posterior hyperparameter values for the model for that examiner. The means of the posterior hyperparameter values from a group of examiners are then used as the prior hyperparameter values for another examiner who was not a member of that group – for the purpose of the demonstration in the present paper, leave-one-out cross-validation is used.

<Figure 1 about here>

**Figure 1.** Outline of proposed method.

In the proposed method, the statistical model used for both the numerator and the denominator of the Bayes factor is a beta-binomial model.[6] Previous uses of beta-binomial models in forensic inference include Cereda [31] in DNA-profile comparison, Rosas et al. [32] in speaker recognition, Song et al. [33] in firearms examination, and Kadane [34] in document examination. For the beta-binomial model, the parameter of interest, $\theta_{(RS|t)}$, is the probability of response $RS$ given the truth $t$, where the subscript $t$ is a placeholder for s or d ($t = \{s, d\}$) The likelihood of an observed response count, $c_{(RS|t)}$, given $n_t$ opportunities for a response to occur, is modelled by the binomial distribution $\text{Bin}\left(c_{(RS|t)} \middle| \theta_{(RS|t)}, n_t\right)$, and the conjugate prior is modelled by the beta distribution $\text{Beta}\left(\theta_{(RS|t)} \middle| a_t, b_t\right)$, in which $a_t$ and $b_t$ are the hyperparameters for the prior distribution. Via Bayes theorem, the posterior distribution of the parameter $\theta_{(RS|t)}$, $\theta^*_{(RS|t)}$, is proportional to the multiplication of the prior distribution and the likelihood, and this simplifies to $\text{Beta}\left(\theta^*_{(RS|t)}\big|a^*_t, b^*_t\right)$, see Equation (2), in which $c_{(\neg RS|t)}$ is the count of responses that are not $c_{(RS|t)}$ given $n_t$ opportunities for a response to occur $\left(c_{(RS|t)} + c_{(\neg RS|t)} = n_t\right)$, and the posterior hyperparameter values are $a^*_t = c_{(RS|t)} + a_t$ and $b^*_t = c_{(\neg RS|t)} + b_t$.

---

[6] For introductions to the beta-binomial model, see Murphy [29] §3.3, and Banks & Tackett [30] §3.2.1.

(2)

$$p\left(\theta^*_{(RS|t)}\big|c_{(RS|t)}, c_{(\neg RS|t)}, a_t, b_t\right)$$

$$\propto \text{Bin}\left(c_{(RS|t)}\big|\theta_{(RS|t)}, n_t\right) \text{Beta}\left(\theta_{(RS|t)}\big|a_t, b_t\right)$$

$$\propto \left(\theta_{(RS|t)}\right)^{c_{(RS|t)}} \left(1 - \theta_{(RS|t)}\right)^{c_{(\neg RS|t)}} \left(\theta_{(RS|t)}\right)^{a_t-1} \left(1 - \theta_{(RS|t)}\right)^{b_t-1}$$

$$\propto \left(\theta_{(RS|t)}\right)^{c_{(RS|t)}+a_t-1} \left(1 - \theta_{(RS|t)}\right)^{c_{(\neg RS|t)}+b_t-1}$$

$$\propto \text{Beta}\left(\theta^*_{(RS|t)}\big|c_{(RS|t)} + a_t, c_{(\neg RS|t)} + b_t\right)$$

$$\propto \text{Beta}\left(\theta^*_{(RS|t)}\big|a^*_t, b^*_t\right)$$

The expected value of the posterior distribution of $\theta^*_{(RS|t)}$ is $\bar{\theta}^*_{(RS|t)}$. This is calculated as in Equation (3), in which $m_t = a_t + b_t$ is the prior pseudo number of fingermark-fingerprint pairs of truth status $t$, and $m^*_t = n_t + m_t = a^*_t + b^*_t$ is the posterior pseudo number of fingermark-fingerprint pairs of truth status $t$.

(3)

$$\bar{\theta}^*_{(RS|t)} = \int_0^1 \theta^*_{(RS|t)} \text{Beta}\left(\theta^*_{(RS|t)} \middle| c_{(RS|t)} + a_t, c_{(\neg RS|t)} + b_t\right) d\theta^*_{(RS|t)}$$

$$= \frac{c_{(RS|t)} + a_t}{c_{(RS|t)} + a_t + c_{(\neg RS|t)} + b_t} = \frac{a_t^*}{n_t + m_t} = \frac{a_t^*}{a_t^* + b_t^*} = \frac{a_t^*}{m_t^*}$$

A Bayes factor, $B_{RS}$, is then calculated as the ratio of the expected values of the posterior distributions of $\theta^*_{(RS|s)}$ and $\theta^*_{(RS|d)}$, as in Equation (4).

(4)

$$B_{RS} = \frac{\bar{\theta}^*_{(RS|s)}}{\bar{\theta}^*_{(RS|d)}} = \frac{a_s^*/m_s^*}{a_d^*/m_d^*}$$

For uninformative priors, the proposed method uses hyperparameters $a_s = b_s = n_s/(n_s+n_d)$ and $a_d = b_d = n_d/(n_s+n_d)$. If $n_s = n_d$, then the hyperparameters equal those for Jeffreys' reference priors: $a = b = 0.5$. If $n_s \neq n_d$, then the priors for the numerator and denominator of the Bayes factor are weighted versions of Jeffreys' reference priors. This weighting prevents the bias that would occur in the calculation of the Bayes factor if unweighted Jeffreys' reference priors were used (see Rosas et al. [32] Appendix A). In the Langenburg et al. [22] data $n_s = 7$ and $n_d = 5$; hence for the Langenburg et al. [22] data $a_s = b_s = n_s/(n_s+n_d) = 7/12$ and $a_d = b_d = n_d/(n_s+n_d) = 5/12$.

In the present paper, for informative priors, a cross-validated procedure was adopted whereby the data from one examiner in a group were held out, the posterior hyperparameter values ($a_s^*$, $b_s^*$, $a_d^*$, and $b_d^*$) for each of the remaining examiners in the group were calculated using uninformative priors as described above, then the means of those posterior hyperparameter values across examiners ($\bar{a}_s^*$, $\bar{b}_s^*$, $\bar{a}_d^*$, and $\bar{b}_d^*$) were calculated,[7] and finally those mean values were used as the hyperparameter values of the informative priors for the held-out examiner's performance. The latter values were substituted as the $a_s$, $b_s$, $a_d$, and $b_d$ values in Equation (3) to calculate the expected values of the posterior parameter distributions for the left-out examiner, which in turn were substituted into Equation (4) to calculate the Bayes factor for the left-out examiner.

---

[7] If the $n_t$ and hence the $m_t^*$ values differed across examiners, a weighted mean could be used.

## 3  Results

Table 2 shows example sample data from one examiner from Group 1. Figure 2 shows a graphical representation of an example of the calculation of a Bayes factor $B_{ID}$ for this examiner using uninformative priors (left panels) and informative priors (right panels). The top panels represent the calculation of the numerators of the Bayes factors, and the bottom panels represent the calculation of the denominators.

**Table 2.** Example sample data consisting of one examiner's counts of "identification", "inconclusive", an "exclusion" outputs in responses to same-source and different-source fingermark-fingerprint pairs.

|  |  | Response | | | Number of pairs |
|---|---|---|---|---|---|
|  |  | identification | inconclusive | exclusion |  |
| Truth | same source | $c_{(ID|s)} = 5$ | $c_{(IN|s)} = 1$ | $c_{(EX|s)} = 1$ | $n_s = 7$ |
|  | different source | $c_{(ID|d)} = 0$ | $c_{(IN|d)} = 1$ | $c_{(EX|d)} = 4$ | $n_d = 5$ |

<Figure 2 about here>

**Figure 2.** Graphical representation of an example of the calculation of a Bayes factor

$B_{ID}$ using uninformative priors, left panels (a) and (b), and informative priors, right panels (c) and (d). Top panels (a) and (c) represent the calculation of numerators, and bottom panels (b) and (d) represent the calculation of denominators. Dotted curves: prior distributions. Dashed vertical lines: sample proportions. Solid curves: posterior distributions. Solid vertical lines: expected values of posterior distributions.

Figure 2(a) represents the calculation of the numerator of the Bayes factor using uninformative priors, including the prior distribution $\text{Beta}\left(\theta_{(ID|s)} \middle| a_s, b_s\right) = \text{Beta}\left(\theta_{(ID|s)} \middle| 7/12, 7/12\right)$ (the dotted curve), the sample proportion $c_{(ID|s)}/n_s = 5/7$ (the dashed vertical line), the posterior distribution $\text{Beta}\left(\theta^*_{(ID|s)} \middle| a_s^*, b_s^*\right) = \text{Beta}\left(\theta^*_{(ID|s)} \middle| 5/7 + 7/12, 2/7 + 7/12\right) = \text{Beta}\left(\theta^*_{(ID|s)} \middle| 5.58, 2.58\right)$ (the solid curve), and the expected value of the posterior distribution $\bar{\theta}^*_{(ID|s)} = a_s^*/m_s^* = 5.58/8.27 = 0.684$ (the solid vertical line). Similarly, Figure 2(b) represents the calculation of the denominator of the Bayes factor using uninformative priors, including the prior distribution $\text{Beta}\left(\theta_{(ID|d)} \middle| a_d, b_d\right) = \text{Beta}\left(\theta_{(ID|d)} \middle| 5/12, 5/12\right)$, the sample proportion $c_{(ID|d)}/n_d = 0/5$, the posterior distribution $\text{Beta}\left(\theta^*_{(ID|d)} \middle| a_d^*, b_d^*\right) = \text{Beta}\left(\theta^*_{(ID|d)} \middle| 0/5 + 7/12, 5/5 + 5/12\right) = \text{Beta}\left(\theta^*_{(ID|d)} \middle| 0.417, 5.42\right)$, and the expected value of the posterior distribution $\bar{\theta}^*_{(ID|d)} = a_d^*/m_d^* = 0.417/5.83 = 0.0714$. The resulting Bayes-factor value is $B_{ID} = \bar{\theta}^*_{(ID|s)} / \bar{\theta}^*_{(ID|d)} = 0.684/0.0714 = 9.57$.

The hyperparameters for the informative priors were calculated as the means of the posterior hyperparameter values for all the other examiners in Group 1, with each of those examiners' posterior hyperparameter values calculated using their response data and uninformative priors. This resulted in informative prior hyperparameter values of $a_s = 4.93$, $b_s = 3.24$, $a_d = 0.591$, and $b_d = 5.24$. Figure 2(c) represents the calculation of the numerator of the Bayes factor using informative priors, including the prior distribution $\text{Beta}(\theta_{(ID|s)}|a_s, b_s) = \text{Beta}(\theta_{(ID|s)}|4.93, 3.24)$, the sample proportion $c_{(ID|s)}/n_s = 5/7$, the posterior distribution $\text{Beta}(\theta^*_{(ID|s)}|a^*_s, b^*_s) = \text{Beta}(\theta^*_{(ID|s)}|5/7 + 4.93, 2/7 + 3.24) = \text{Beta}(\theta^*_{(ID|s)}|9.93, 5.24)$, and the expected value of the posterior distribution $\bar{\theta}^*_{(ID|s)} = a^*_s/m^*_s = 9.93/15.2 = 0.655$. Similarly, Figure 2(d) represents the calculation of the denominator of the Bayes factor, including the prior distribution $\text{Beta}(\theta_{(ID|d)}|a_d, b_d) = \text{Beta}(\theta_{(ID|d)}|0.591, 5.24)$, the sample proportion $c_{(ID|d)}/n_d = 0/5$, the posterior distribution $\text{Beta}(\theta^*_{(ID|d)}|a^*_d, b^*_d) = \text{Beta}(\theta^*_{(ID|d)}|0/5 + 0.591, 5/5 + 5.24) = \text{Beta}(\theta^*_{(ID|d)}|0.591, 10.2)$, and the expected value of the posterior distribution $\bar{\theta}^*_{(ID|d)} = a^*_d/m^*_d = 0.591/10.8 = 0.0545$. The resulting Bayes-factor value is $B_{ID} = \bar{\theta}^*_{(ID|s)}/\bar{\theta}^*_{(ID|d)} = 0.655/0.0545 = 12.0$.

For this example, as detailed above, the $B_{ID}$ values calculated using uninformative priors and informative priors were 9.57 and 12.0 respectively. Figure 3 shows a graphical representation of an example of the calculation of $B_{IN}$ for the same examiner. The $B_{IN}$ values calculated using uninformative priors and informative priors were 1/1.25 and 1.22 respectively. Figure 4 shows a graphical representation of an example of the calculation of $B_{EX}$ for the same examiner. The $B_{EX}$ values calculated using uninformative priors and informative priors were 1/3.91 and 1/5.75 respectively.[8]

---

[8] By convention, values have been reported to 3 significant figures. Given the small data set, the resolution of the Bayes-factor values is probably not meaningful past 1 significant figure.

<Figure 3 about here>

**Figure 3.** Graphical representation of an example of the calculation of a Bayes factor $B_{IN}$ using uninformative priors, left panels (a) and (b), and informative priors, right panels (c) and (d). Top panels (a) and (c) represent the calculation of numerators, and bottom panels (b) and (d) represent the calculation of denominators. Dotted curves: prior distributions. Dashed vertical lines: sample proportions. Solid curves: posterior distributions. Solid vertical lines: expected values of posterior distributions.

<Figure 4 about here>

**Figure 4.** Graphical representation of an example of the calculation of a Bayes factor $B_{EX}$ using uninformative priors, left panels (a) and (b), and informative priors, right panels (c) and (d). Top panels (a) and (c) represent the calculation of numerators, and bottom panels (b) and (d) represent the calculation of denominators. Dotted curves: prior distributions. Dashed vertical lines: sample proportions. Solid curves: posterior distributions. Solid vertical lines: expected values of posterior distributions.

Figure 5 shows Bayes-factor values calculated for each examiner in Group 1. The left panel, panel (a), shows the Bayes-factor values calculated using uninformative priors and the right panel, panel (b), shows the Bayes-factor values calculated using informative priors. Bayes-factor values are plotted using a base-2 logarithmic scale: A $\log_2$ Bayes-factor value of +1 is a Bayes-factor value of 2, a $\log_2$ Bayes-factor value of +2 is a Bayes-factor value of 4, $\log_2$ Bayes-factor value of +3 is a Bayes-factor value of 8, etc., and a $\log_2$ Bayes-factor value of −1 is a Bayes-factor value of 1/2, a $\log_2$ Bayes-factor value of −2 is a Bayes-factor value of 1/4, $\log_2$ Bayes-factor value of −3 is a Bayes-factor value of 1/8, etc. A $\log_2$ Bayes-factor value of 0 is a Bayes-factor value of 1.

<Figure 5 about here>

**Figure 5.** Swarm chart of Bayes-factor values for each examiner in Group 1. (a) using uninformative priors. (b) using informative priors.

Compared to using uninformative priors, using informative priors resulted in tighter grouping of examiners' Bayes-factor values for each of $B_{ID}$, $B_{IN}$, and $B_{EX}$. Also, on average across examiners, using informative priors resulted in larger $B_{ID}$ values and smaller $B_{EX}$ values.

Using informative priors, "identification" responses converted to relatively large $B_{ID}$ values in favour of the same-source hypothesis, "inconclusive" responses converted to relatively small $B_{IN}$ values in favour of the same-source hypothesis, and "exclusion" responses converted to relatively large $B_{EX}$ values in favour of the different-source hypothesis. Note that, for a substantial proportion of examiners, "inconclusive" responses did not correspond to a neutral strength of evidence, they did not result in $B_{IN}$ values of approximately 1, they resulted in $B_{IN}$ values somewhat above 1.

The maximum and minimum Bayes-factor values were constrained by the number of fingermark-fingerprint pairs.

Using uninformative priors, the largest $B_{ID}$ value obtained was 13 and the smallest $B_{EX}$ value obtained was 1/13. These are the maximum and minimum values that could be obtained using 12 fingermark-fingerprint pairs. If an examiner had both $B_{ID} = 13$ and $B_{EX} = 1/13$, this was the result of perfect responses, i.e., "identification" in response to all same-source pairs, "exclusion" in response to all different-source pairs, and no "inconclusive" responses. In general, the largest Bayes factor value that could be obtained using this method with uninformative priors would be $n_s + n_d + 1$ and the smallest would be $1/(n_s + n_d + 1)$.

Using informative priors, the largest $B_{ID}$ value obtained was 13.2 and the smallest

$B_{EX}$ value obtained was 1/11.9. Theoretically, given the Langenburg et al. (2012) data, the largest Bayes factor value that could have been obtained using informative priors would have been $2(n_s + n_d) + 1 = 25$, and the smallest would have been $1/(2(n_s + n_d) + 1) = 1/25$, but this would have required perfect responses from all participants.

## 4  Discussion

The present paper has proposed and demonstrated a method to convert traditional fingerprint-examination conclusions to well-calibrated Bayes factors. The method requires minimal changes to existing practice. Examiners continue to initially state their ACE or ACE-V outputs as "identification", "inconclusive", and "exclusion", and a statistical model is then used to calculate the strength of evidence associated with each of these outputs.

The demonstration used a convenient dataset of individual examiners' ACE responses to each of multiple fingermark-fingerprint pairs. In the context of a case, the system which would have to be calibrated would be the system which is actually used to calculate the strength of evidence associated with the questioned-source fingermark and known-source fingerprint of interest in the case. If a system consisted of an implementation of the ACE-V process by a particular primary examiner and a particular secondary examiner, then that is the system that would have to be calibrated. If the process used for casework involved comparison of multiple candidate fingerprints with a fingermark (rather than a single print with a single mark), then that would form part of the system that would have to be calibrated. If the processes used for casework involved consideration by examiners of the scores output by an automatic fingerprint identification system (AFIS), then that would form part of the system that would have to be calibrated.

The system would have to be calibrated under conditions reflecting the conditions of the case under consideration. In order for the calculated Bayes-factor value to be meaningful, the system would have to provide responses to fingermark-fingerprint pairs for which the true same-source or different-source status is known and which are sufficiently representative of the relevant population and sufficiently reflective of the conditions of the case under consideration. Those responses could then be used to train the statistical model. Decisions about whether fingermark-fingerprint pairs are sufficiently representative of the relevant population and sufficiently reflective of the conditions of the case under consideration require subjective judgement based on subject-area expertise (Morrison et al. [23]). A key consideration will be the quality of the fingermark. Ideally, examiners (and researchers with subject-area expertise) would collaboratively define a limited number of commonly-encountered sets of conditions, pairs of marks and prints reflecting each of those sets of conditions would be created, and each system would provide responses to pairs from each set of conditions. In a casework context, an examiner (or process involving multiple examiners) would assess whether the fingermark-fingerprint pair was sufficiently similar to one of the sets of conditions for which a model already exists, and, if so, would select the appropriate model to use for the case. The examiner would then use the selected model to convert the system's "identification", "inconclusive", or "exclusion" output to the corresponding Bayes-factor value. In casework, this conversion simply requires looking up the selected model's Bayes-factor value corresponding to the chosen categorical output. For a given system, examiners could be provided with a table of conversion values for each output under each set of conditions.

The demonstration used a convenient dataset of individual examiners' responses to each of only 12 fingermark-fingerprint pairs. This resulted in constrained Bayes-factor values, i.e., the maximum and minimum Bayes-factor values achievable could not be very far from 1. This reflects the desired behaviour of a method for calculating Bayes factors: To make stronger strength-of-evidence claims, one would need more evidence to support those claims in the form of more correct responses to fingermark-fingerprint pairs, which would require more opportunities to give correct responses to fingermark-fingerprint pairs. When the number of opportunities to give correct responses to fingermark-fingerprint pairs is small, the strength-of-evidence claims that can potentially be supported are weaker. In order to be able to potentially make stronger strength-of-evidence claims in casework, the system to be calibrated would have to provide responses to a large number of fingermark-fingerprint pairs for which the true same-source or different-source status was known. This would have to be repeated for each set of conditions for which one wanted to potentially make stronger strength-of-evidence claims.

An advantage of the beta-binomial model is that training data do not have to be provided all at once. Each time a response to a new pair is provided, the model can be updated. A large number of responses could therefore be built up over a long period of time. To train an initial model for a system under a set of conditions, one might initially present the system with a relatively large number of pairs, but thereafter one could institute periodic presentation of smaller numbers of pairs, or could present an ongoing trickle of pairs. If a laboratory were using a quality-management process which included blind testing, i.e., inserting tests into examiners' workflows in such a way that examiners do not know that they are tests, the system's response to each such test could be used to update the model. Over time, using periodic or trickle testing, the model would better represent the system's performance and would potentially support stronger strength-of-evidence claims. If the system's performance changed over time, periodic or trickle testing would provide the data necessary to update the model to reflect that change.

If examiners wanted to adopt an ordinal scale with more than three levels, or wanted to adopt subjective assignment of continuous likelihood-ratio values, those ordinal or continuous values could be calibrated using other statistical models. A commonly used model for calibrating continuously-valued likelihood ratios (that can also be applied to ordinal scales) is logistic regression (Brümmer & du Preez [35]; González-Rodríguez et al. [36]; Morrison [37], [38]; Morrison & Poh [39]).

The introduction of a method such as that proposed in the present paper could potentially lead to a gradual change in practice. Examiners could potentially use the method to inform their practice, e.g., feedback in the form of the Bayes-factor value associated with each of the categorical outputs ("identification", "inconclusive", "exclusion") in each set of conditions could lead to examiners adjusting where they set the categorical boundaries dependent upon the conditions. Examiners exposed to the proposed method could also become accustomed to probabilistic reasoning and in the future could be more willing to accept other probabilistic methods as useful tools to assist them in assessing strength of evidence.

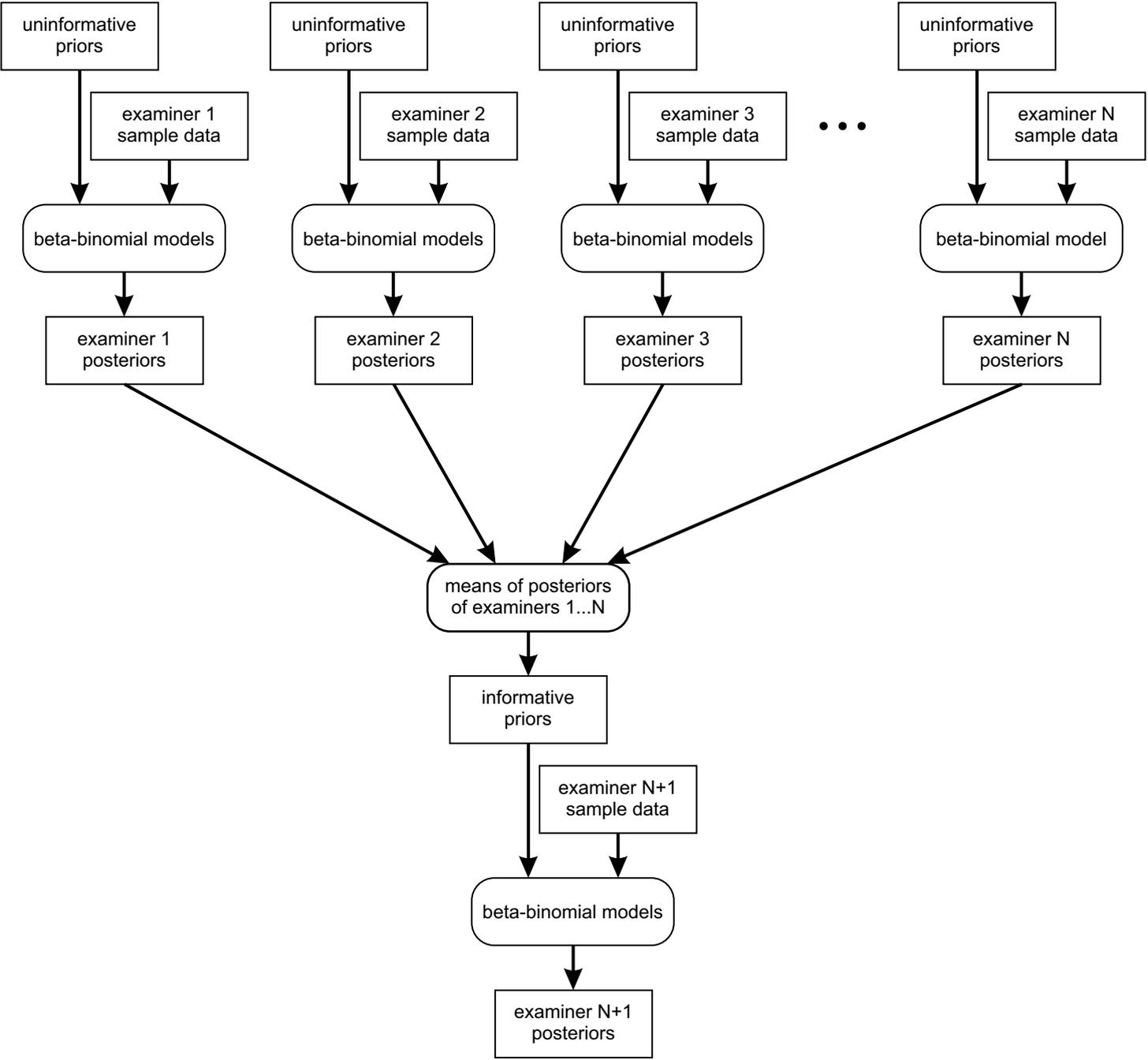

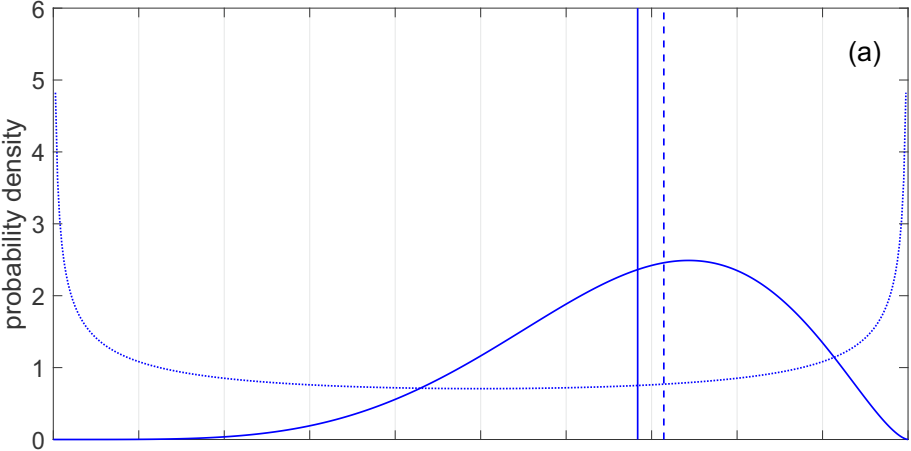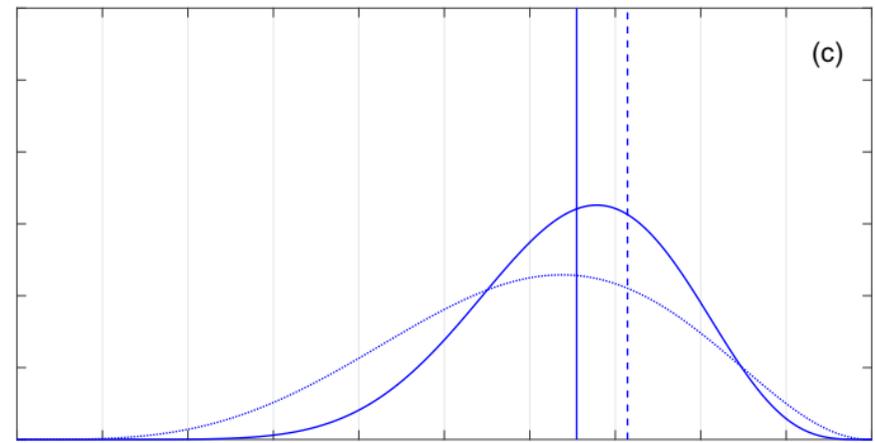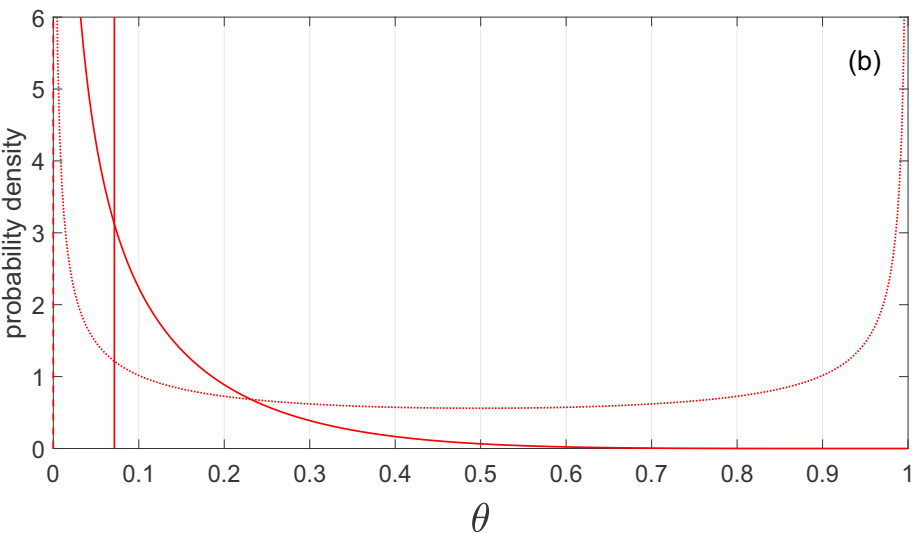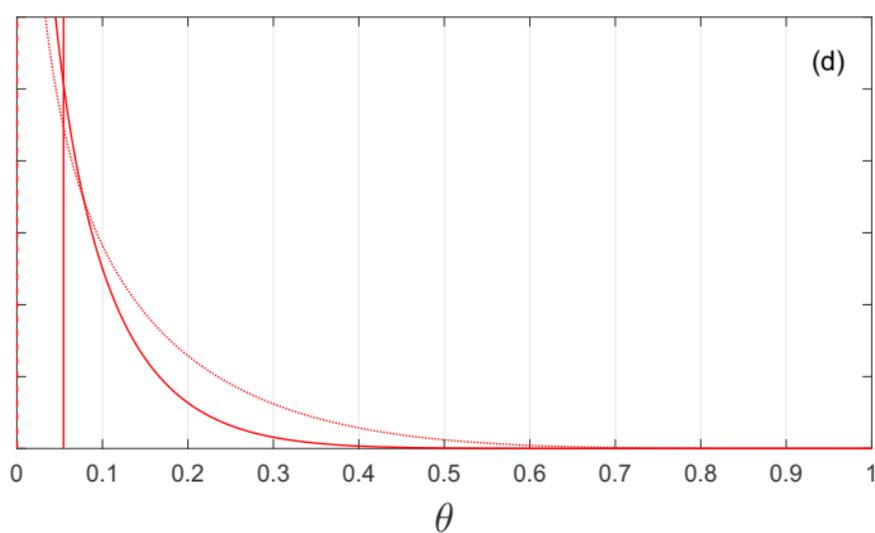

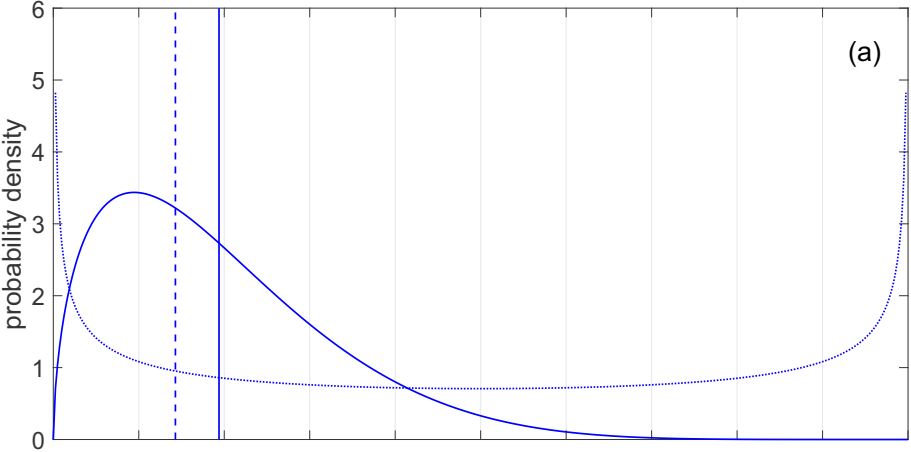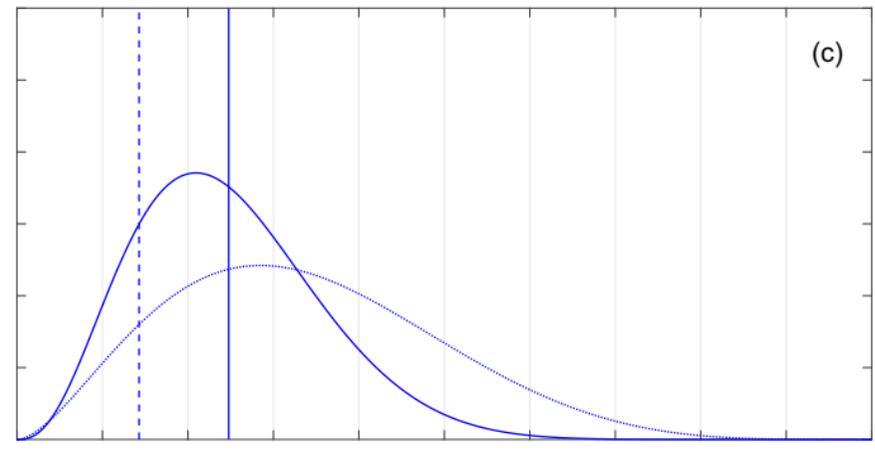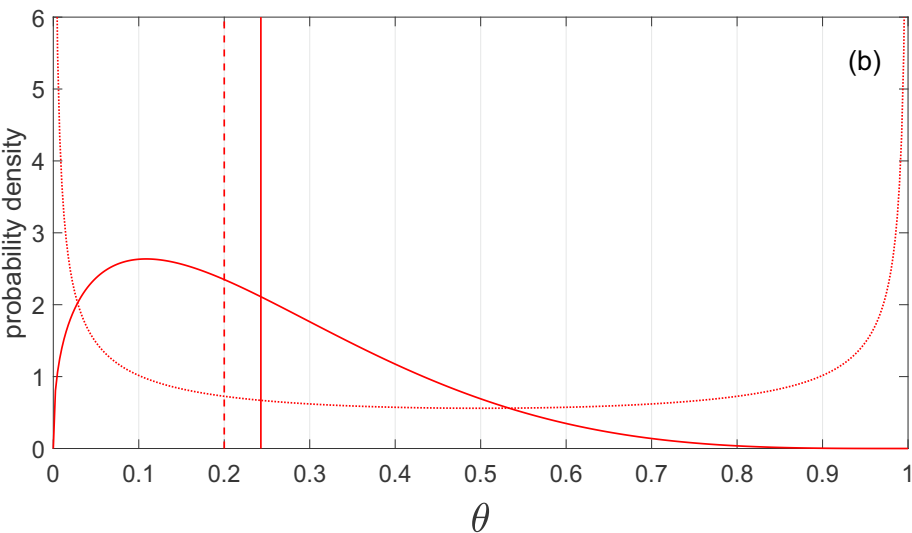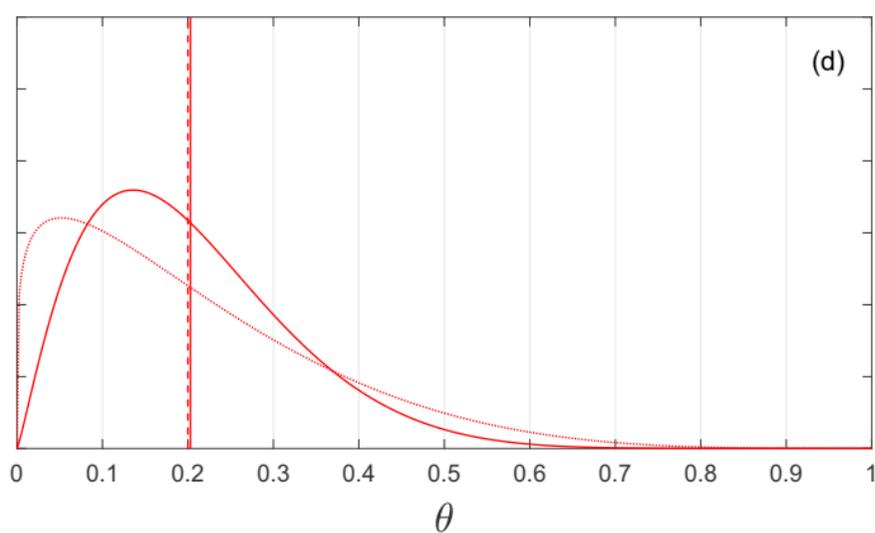

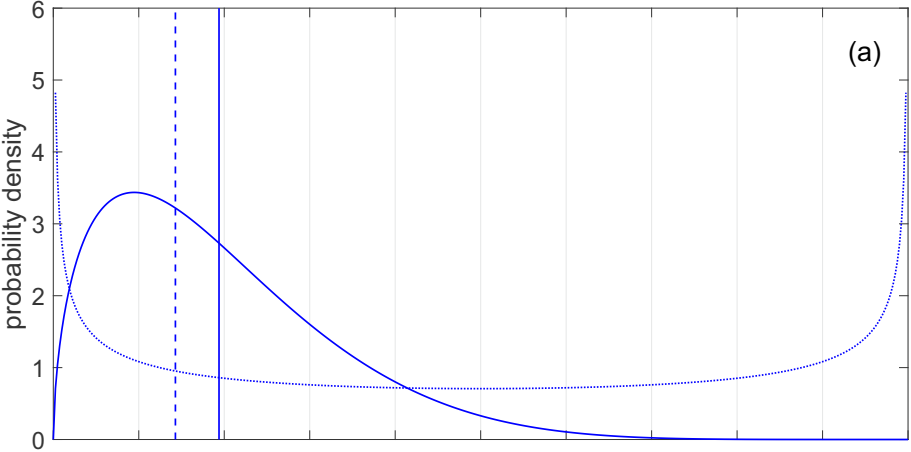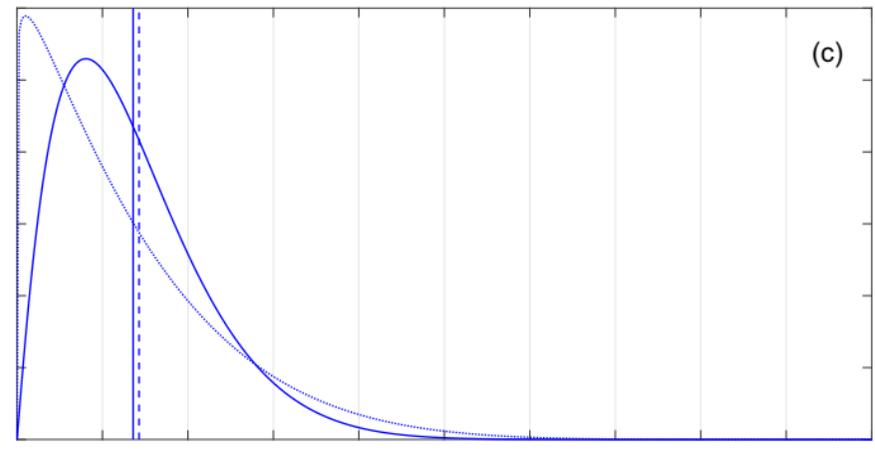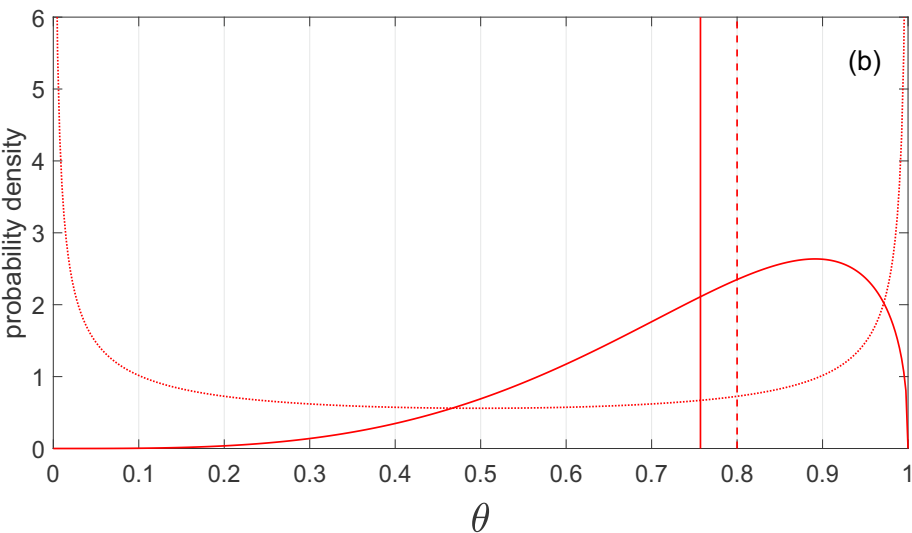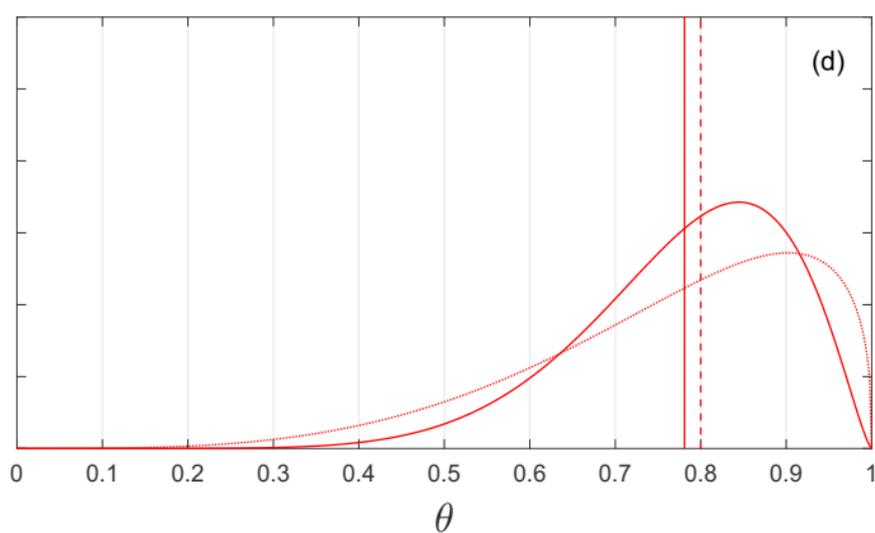

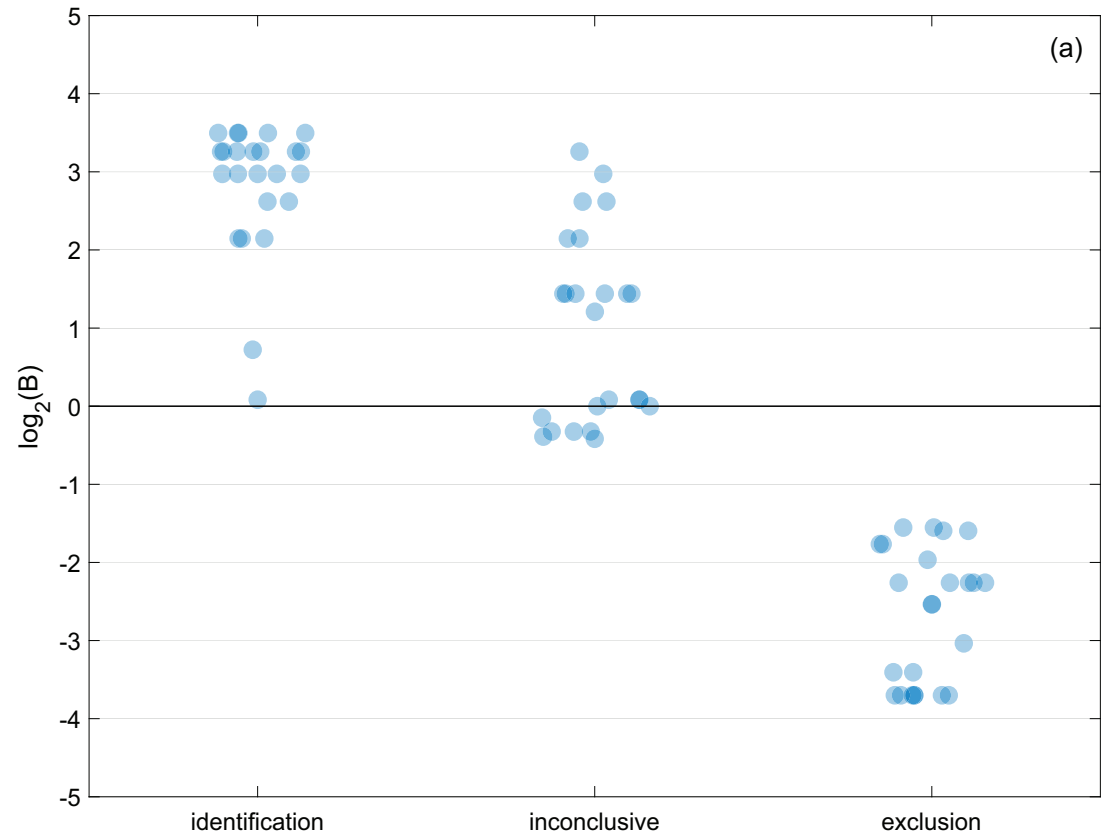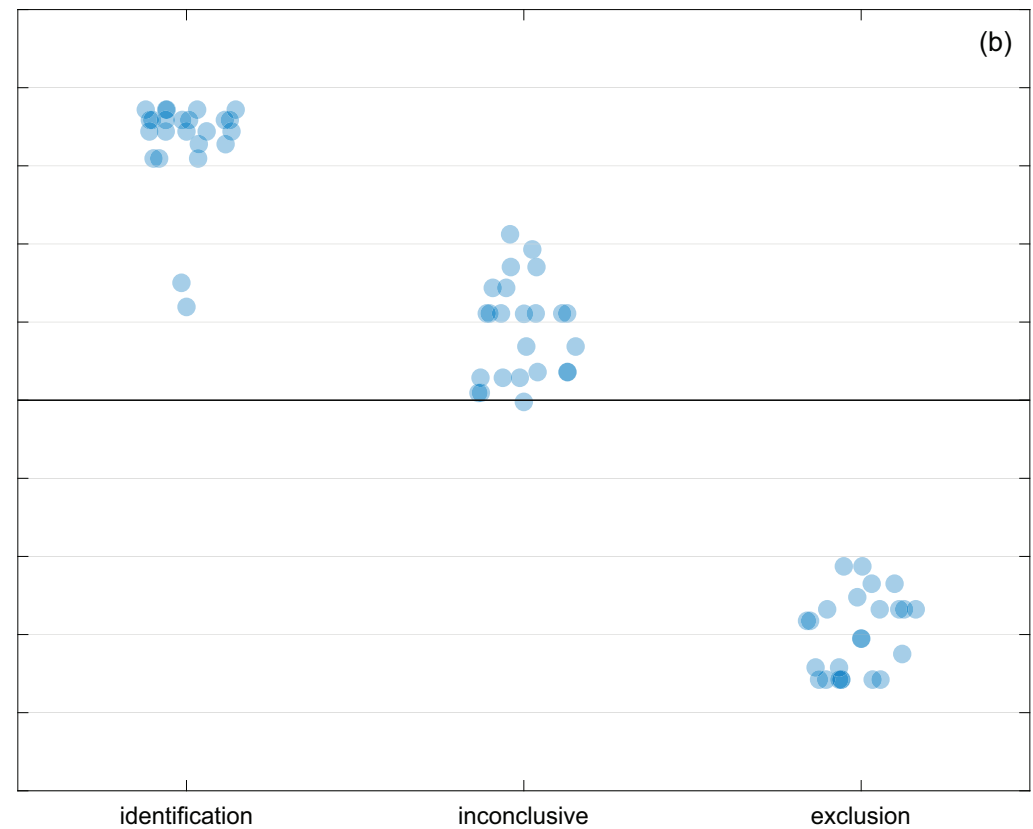